\documentstyle[psfig,aps]{revtex}
\def\ts{\textstyle}
\begin{document}
\title{An extension of the Lyapunov analysis for the predictability
problem}

\author{G. Boffetta}
\address{Dipartimento di Fisica Generale,
Universit\`a di Torino, via Pietro Giuria 1, 10125 Torino, Italy}

\author{P. Giuliani, G. Paladin}
\address{Dipartimento di Fisica, Universit\`a dell'Aquila,
         via Vetoio, Coppito 67100 L'Aquila, Italy}

\author{A. Vulpiani}
\address{Dipartimento di Fisica, Universit\`a di Roma ``La Sapienza'',
         p.le Aldo Moro 2, 00185 Roma, Italy}

\date{\today}

\maketitle

\begin{abstract}
The predictability problem for systems with different characteristic time
scales is investigated. It is shown that even in simple
chaotic dynamical systems, the leading Lyapunov
exponent is not sufficient to estimate the predictability time.
This fact is due the saturation of the error on the fast
components of the system which therefore do not contribute to the 
exponential growth of the error at large errors.
It is proposed to adopt a
generalization of the Lyapunov exponent which is based on the natural
concept of error growing time at finite error size. The method is
first illustrated on a simple numerical model obtained by coupling two
Lorenz systems with different time scales.
As a more realistic example, this analysis
is then applied to a toy model of Atmospheric circulation recently
introduced by Lorenz.
\end{abstract}

\section{Introduction}
\label{sec:1}
The prediction of the future state of a system known the actual state is
a fundamental problem with obvious applications in geophysical flows
(Leith, 1975; Leith and Kraichnan, 1972; Leith, 1978).
There are many limitations to the ability of predicting the state of a
geophysical system, e.g. the atmosphere, one of the most important
is the lack of knowledge, or the difficulty of full implementation, of the 
equations of motion. Still, even if one
assume to perfectly know the system and to have sufficiently
large computers, the predictability can be severely limited by the
dynamics itself, i.e. the ``intrinsic unpredictability'' the a system which 
is the subject of our study.

A well known, and very popular, example of low-predictable system is
given by a chaotic system
(Lorenz, 1963). 
By definition, chaotic dynamical systems display
sensible dependence on initial conditions: two initially
close trajectories will diverge exponentially in the phase space with a
rate given by the leading Lyapunov exponent $\lambda_{max}$ 
(see Eckmann and Ruelle, 1985).
Because the initial
condition can be measured only with a finite uncertainty $\delta$, 
we can know the future state of the system at a tolerance level
$\Delta$ only up to a maximum time
\begin{equation}
T_{p} \sim {1\over \lambda_{max}} \, \ln\left({\Delta \over
                                         \delta}\right)
\label{eq:1}
\end{equation}
One important consequence of equation (\ref{eq:1}) is that the
predictability time has a very weak dependence on the precision of the
initial condition and on the tolerance, therefore the predictability
time is an intrinsic quantity of the system as the Lyapunov exponent
is.

The naive formula (\ref{eq:1}) for the predictability problem 
holds only for infinitesimal perturbations and in non intermittent
systems; in the general case one has a series of problems and subtle 
points which have been objected of several studies in last years 
(Crisanti et al., 1993; Aurell et al., 1996, 1997).
One delicate issue is particularly relevant for our present study and 
essentially says that,
although the Lyapunov exponent for the atmosphere (as a whole) is
presumably rather large (due to the small scale turbulence), the large
scale behavior of the system can be forecasted with good accuracy for 
several days 
(Lorenz, 1969; Lorenz, 1982; Simmons et al., 1995).

The apparent paradox comes from the identification of the 
predictability time with
the inverse of the Lyapunov exponent based on equation (\ref{eq:1}) which is 
actually of little relevance even in few degree-of-freedom dynamical
systems. Indeed, in presence of different characteristic time scales, as
is the case in any realistic model of geophysical flows, the Lyapunov 
exponent will
be roughly proportional to the inverse smallest characteristic time.
This time is associated to the smallest, low energy containing scales
which, after the fast saturation, do not play a role any more in the
error growth law. Large errors will grow, in general, with the
characteristic time of the largest, energy containing scales
(Leith, 1971; Leith and Kraichnan, 1972).
Thus when the initial error is not very small, as is often the case in
a predictability experiment, the leading Lyapunov exponent may play no
role at all.

To be more quantitative, in this paper we investigate the predictability 
problem in two time scale dynamical systems. We apply a recently
introduced generalization of the Lyapunov exponent to finite
perturbations. We will show that the Finite Size Lyapunov Exponent
(FSLE) is more suitable for characterizing the predictability of complex
systems where the growth rate of large errors in not ruled by the
Lyapunov exponent.

The models considered here are crude approximations of a realistic
geophysical flow also
because both the subsystems have a single time scale. It would 
be interesting to extend the investigation to more realistic 
situations and comparing the latter case with present results.

This remaining of the paper is organized as follows: in section 
\ref{sec:2} we introduce the Finite Size Lyapunov Exponent which is
applied to the system models in section \ref{sec:3}. Section \ref{sec:4}
is devoted to conclusions.

\section{The Finite Size Lyapunov Exponent}
\label{sec:2}

The notion of Lyapunov exponent is based on the average rate of
exponential separation of two infinitesimally close trajectory in 
the phase space:
\begin{equation}
\lambda_{max} = \lim_{t \rightarrow \infty} \lim_{\delta x(0) \rightarrow 0}
{1 \over t} \ln {\delta x(t) \over \delta x(0)}
\label{eq:b1}
\end{equation}
where $\delta x(t)$ is the distance between the trajectories with a
suitable norm and the two limits cannot be interchanged. The standard
algorithm 
(Benettin et al., 1980) 
for computing the Lyapunov exponent is based 
on (\ref{eq:b1}), with the trick of periodical rescaling of the two 
trajectory in order to keep their distance ``infinitesimal''. 

As already discussed in the previous section, the second limit in
(\ref{eq:b1}) is of dubious interest in the predictability problem 
because the initial incertitude on the system variables is in general
not infinitesimal. Therefore one would like to relax the infinitesimal
constrain still maintaining some well defined
mathematical properties. Recently, a generalization of
(\ref{eq:b1}) which allows to compute the average exponential separation 
of two trajectories at finite errors $\delta$ have been introduced.
The Finite Size Lyapunov Exponent, $\lambda(\delta)$, 
is based on the concept of error growing
time $T_{r}(\delta)$ which is the time it takes for a perturbation of
initial size $\delta$ to grow of a factor $r$. The ratio $r$ should not
be taken too large, in order to avoid the growth through different
scales. The error growing time is a fluctuating quantity and one has to
take the average along the trajectory as in (\ref{eq:b1}). The Finite Size
Lyapunov Exponent is then defined as
\begin{equation}
\lambda(\delta) = 
\left\langle {1 \over T_{r}(\delta)} \right\rangle_{t} \ln r =
{1 \over \langle T_{r}(\delta) \rangle} \ln r
\label{eq:b2}
\end{equation}
where $\langle ... \rangle_{t}$ denotes the natural measure along the
trajectory and $\langle ... \rangle$ is the average over many
realizations. For an exhaustive discussion on the way to take averages,
see 
Aurell et al. (1997).

In the limit of infinitesimal perturbations, $\delta \rightarrow 0$,
definition (\ref{eq:b2}) reduces to that of the leading Lyapunov exponent
(\ref{eq:b1}). In practice, $\lambda(\delta)$ displays a plateau at the value
$\lambda_{max}$ for sufficiently small $\delta$.
 
To practically compute the FSLE, one has first
to define a series of thresholds $\delta_{n}=r^{n} \delta_{0}$, and to
measure the time $T_r(\delta_{n})$ that a perturbation with
size $\delta_{n}$ takes to grow up to $\delta_{n+1}$. The time
$T_r(\delta_{n})$ is obtained by following the evolution of the
perturbation from its initial
size $\delta_{\rm min}$ up to the largest threshold $\delta_{\rm max}$.
This is done by integrating two trajectories of the
system that start at an initial distance $\delta_{\rm min}$.
In general, one must take $\delta_{\rm min}\ll\delta_{0}$,
in order to allow the direction of the initial perturbation
to align with the most unstable phase-space direction.
The FSLE, $\lambda(\delta_{n})$, is then computed by averaging the
error growing times over several realizations according to
(\ref{eq:b2}).

Note that the  FSLE has conceptual similarities with
the $\epsilon$-entropy 
(Kolmogorov, 1956; see also Gaspard and Wang, 1993). 
This latter measures
the bandwidth that is necessary for reproducing the trajectory of
a system within a finite accuracy $\delta$. The $\epsilon$-entropy
approach has already been applied to the analysis of simple systems 
and experimental data 
(Gaspard and Wang, 1993), 
giving interesting results. 
The calculation of the $\epsilon$-entropy is, however, much more
expensive from a computational point of view and of little relevance for
the predictability problem.

The computation of the FSLE gives information on the typical
predictability time for a trajectory with initial incertitude $\delta$.
To be more quantitative, one can introduce the average predictability
time from an initial error $\delta$ to a given tolerance $\Delta$ as
the average error growing time, i.e.
\begin{equation}
T_{p}=\int_{\delta}^{\Delta} {d \ln \delta' \over \lambda(\delta')}
\label{eq:b3}
\end{equation}
which reduces to (\ref{eq:1}) in the case of constant $\lambda$.
From general considerations, one expects that $\lambda(\delta)$ is a
decreasing function of $\delta$ and thus (\ref{eq:b3}) gives longer
predictability time than (\ref{eq:1}). 

\section{The models}
\label{sec:3}

We now discuss the application of the FSLE analysis to two relatively 
simple dynamical systems presenting different characteristic time
scales. The proposed models are of little physical relevance; they 
should rather be intended as prototypical models for the predictability
problem in complex flows. 

The first example is obtained by coupling two Lorenz models
(Lorenz, 1963),
the first representing the slow dynamics and the second the fast
dynamics
\begin{equation}
\left\{
\begin{array}{lll}
{\ts d x^{(s)}_{1} \over \ts d t} & = 
  & \sigma (x^{(s)}_{2} - x^{(s)}_{1}) \\
{\ts d x^{(s)}_{2} \over \ts d t} & = 
  & (- x^{(s)}_{1} x^{(s)}_{3} + r_s x^{(s)}_{1} - 
  x^{(s)}_{2}) - \epsilon_s x^{(f)}_{1} x^{(f)}_{2} \\
{\ts d x^{(s)}_{3} \over \ts d t} & = 
  & x^{(s)}_{1} x^{(s)}_{2} - b x^{(s)}_{3} \\
{\ts d x^{(f)}_{1} \over \ts d t} & = 
  & c \sigma (x^{(f)}_{2} - x^{(f)}_{1}) \\
{\ts d x^{(f)}_{2} \over \ts d t} & = 
  & c (- x^{(f)}_{1} x^{(f)}_{3} + r_f x^{(f)}_{1} -
  x^{(f)}_{2}) + \epsilon_f x^{(f)}_{1} x^{(s)}_{2} \\
{\ts d x^{(f)}_{3} \over \ts d t} & = 
  & c (x^{(f)}_{1} x^{(f)}_{2} - b x^{(f)}_{3}) 
\end{array}
\right.
\label{eq:c1} 
\end{equation}

The choice of the form of the coupling is constrained by the physical 
request that the
solution remains in a bounded region of the phase space. Since
\begin{equation}
\begin{array}{l}
{\ts d \over \ts d t} \left\{ 
\epsilon_{f} \left( {(x^{(f)}_{1})^2 \over 2 \sigma} + 
{(x^{(f)}_{2})^2 \over 2} + {(x^{(f)}_{3})^2 \over 2} - 
(r_{f}+1) x^{(f)}_{3} \right) + \right. \\
\\
\left. \epsilon_{s} \left( {(x^{(s)}_{1})^2 \over 2 \sigma} + 
{(x^{(s)}_{2})^2 \over 2} + {(x^{(s)}_{3})^2 \over 2} - 
(r_{s}+1) x^{(s)}_{3} \right) \right\} < 0 ,
\end{array}
\label{eq:c2}
\end{equation}
if the trajectory is enough far from the origin, one has that it
evolves in a bounded region of the phase space.
The parameters have the values $\sigma=10$, $b=8/3$ and $c=10$, the 
latter giving the relative time scale between the fast and slow
dynamics. The two Rayleigh numbers are taken different, $r_s=28$ and
$r_f=45$, in order to avoid sincronization effects.

With the present choice, the two uncoupled systems 
($\epsilon_s=\epsilon_f=0$) display chaotic dynamics with Lyapunov exponent
$\lambda^{(f)} \simeq 12.17$ and $\lambda^{(s)} \simeq 0.905$ respectively 
and thus a relative intrinsic time scale of order $10$.

\begin{figure}
\centerline{\psfig{file=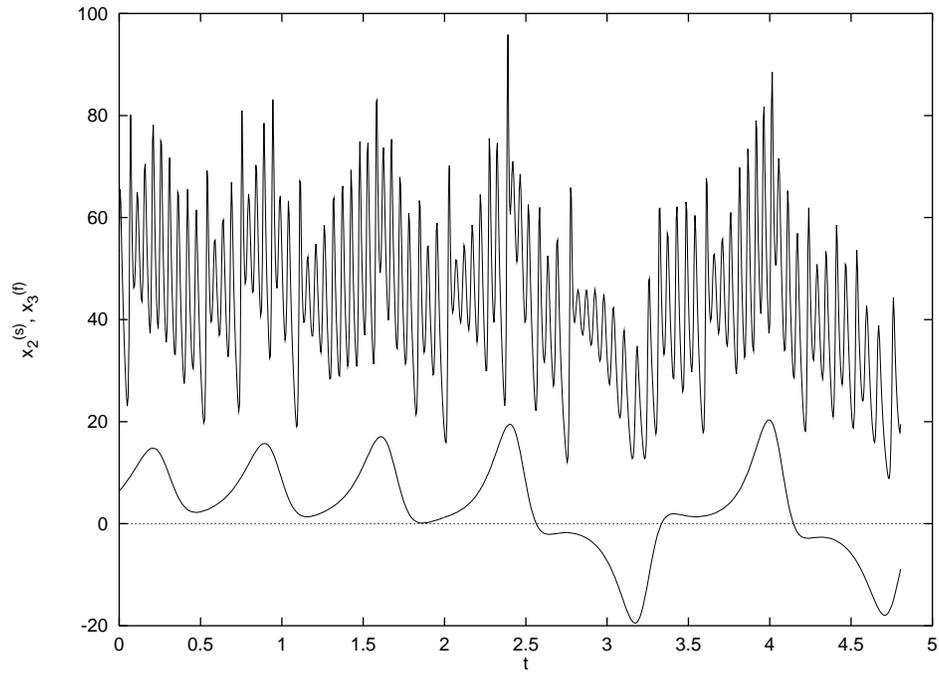}}
\caption{Time series of the slow variable $y_{s}$ (lower curve) and 
of the fast variable $z_{f}$ (upper curve) on the attractor.
}
\label{fig1}
\end{figure}

By switching on the couplings $\epsilon_s$ and $\epsilon_f$ we obtain
a single dynamical system whose maximal Lyapunov exponent $\lambda_{max}$
is close (for small couplings) to the Lyapunov exponent of the faster
decoupled system ($\lambda^{(f)}$). We will consider a single realization
of the couplings, with $\epsilon_f=10$ and $\epsilon_s=10^{-2}$.
The global Lyapunov exponent is found to be in this case 
$\lambda_{max} \simeq 11.5$ which is indeed close to $\lambda^{(f)}$ 
in the uncoupled case.
With the present choice of the couplings, the fast dynamics is driven by 
means of the effective Rayleigh number 
$r_{eff} = r_{f} + \epsilon_{f} x^{(s)}_{2}(t)/c$ and one recognize in
the time evolution
the slow-varying component of the driver (see figure \ref{fig1}).

For what concern the predictability, one expect reasonably that for
small coupling $\epsilon_s$ the slow component of the system ${\bbox
x}_{s}$ remains predictable up to its own characteristic time. On the
other hand, for any coupling $\epsilon \neq 0$ we obtain a single
dynamical system in which the errors grow with the leading Lyapunov
exponent $\lambda_{max} \simeq \lambda^{(f)}$. The apparent paradox stems from
saturation effects which becomes apparent as soon as one is interested
in non infinitesimal errors.

We have integrated two trajectories of (\ref{eq:c1}) starting from very
close initial conditions. One trajectory
represents the ``true'' (reference) trajectory 
${\bbox x}$ and the other is the forecast (perturbed trajectory ${\bbox x}'$) 
subjected to an initial error $\delta x(0)$.
The error is computed here by means of the Euclidean distance in the 
phase space
\begin{equation}
\delta x(t) = \left( \delta x_{f}(t)^2 + \delta x_{s}(t)^2 \right)^{1/2} =
\left[ \sum_{i=1}^{3} \left(x'^{(f)}_{i} - x^{(f)}_{i}\right)^2 + 
\sum_{i=1}^{3} \left(x'^{(s)}_{i} - x^{(s)}_{i}\right)^2 \right]^{1/2}
\label{eq:c3}
\end{equation}

\begin{figure}
\centerline{\psfig{file=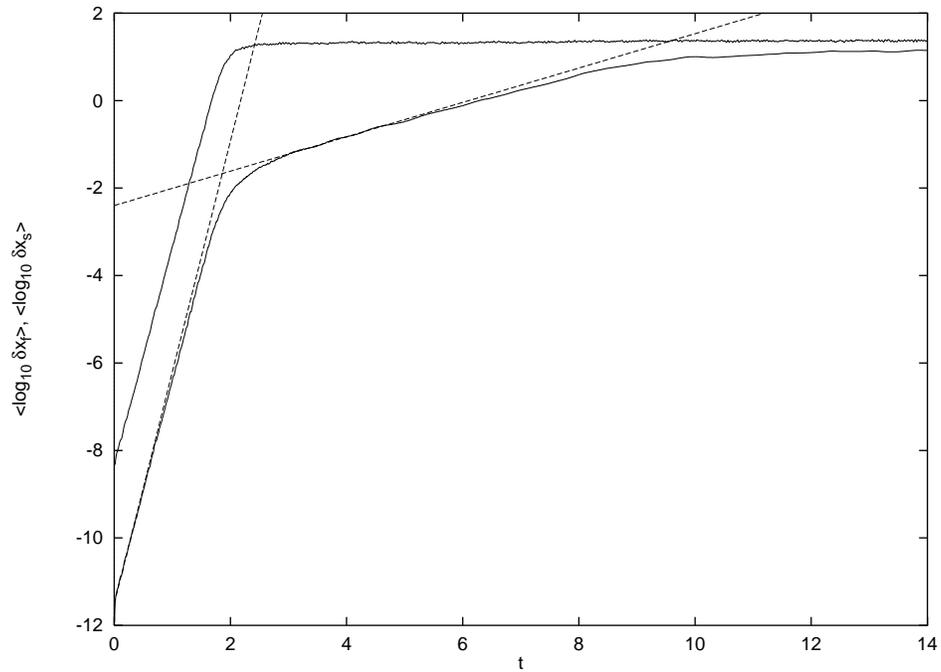}}
\caption{Typical error growth for the fast component $\delta x_{f}$ 
(upper curve) and for the slow component $\delta x_{s}$ in the coupled 
Lorenz models with $\delta x_{f}(0)=10^{-8}$ and 
$\delta x_{s}(0)=10^{-12}$, averaged over $500$ samples. 
In order to detect the typical behavior we compute the average of the
logarithm. The dashed lines show the exponential growths with 
exponents $\lambda^{(f)}$ and $\lambda^{(s)}$.
}
\label{fig2}
\end{figure}

Figure \ref{fig2} reports the results for the error growth 
averaged over $500$ experiments with 
$\delta x_{f}(0)=10^{-8}$ and $\delta x_{s}(0)=10^{-12}$. 
We observe that the relative magnitude of the initial errors is 
irrelevant for what concerns small errors because the error direction 
in the phase space will be rapidly aligned toward the most unstable 
direction.
For small times ($t \leq 2$), both the errors can be considered
infinitesimal and the growth rate is thus given by the global Lyapunov 
exponent $\lambda_{max}$. This is the linear regime of the error growth
in which the Lyapunov exponent is the relevant parameter for the
predictability.
For larger times, the fast component of the error, $\delta x_{f}$, reaches
the saturation, the trajectories separation evolves according to the full
non linear equations of motion and the growth rate for the slow component 
is strongly reduced. From figure \ref{fig2} one observes that the slow 
component error $\delta x_{s}$ is still well below the saturation value,
and grows with a rate close to its characteristic inverse time 
$\lambda^{(s)}$.

\begin{figure}
\centerline{\psfig{file=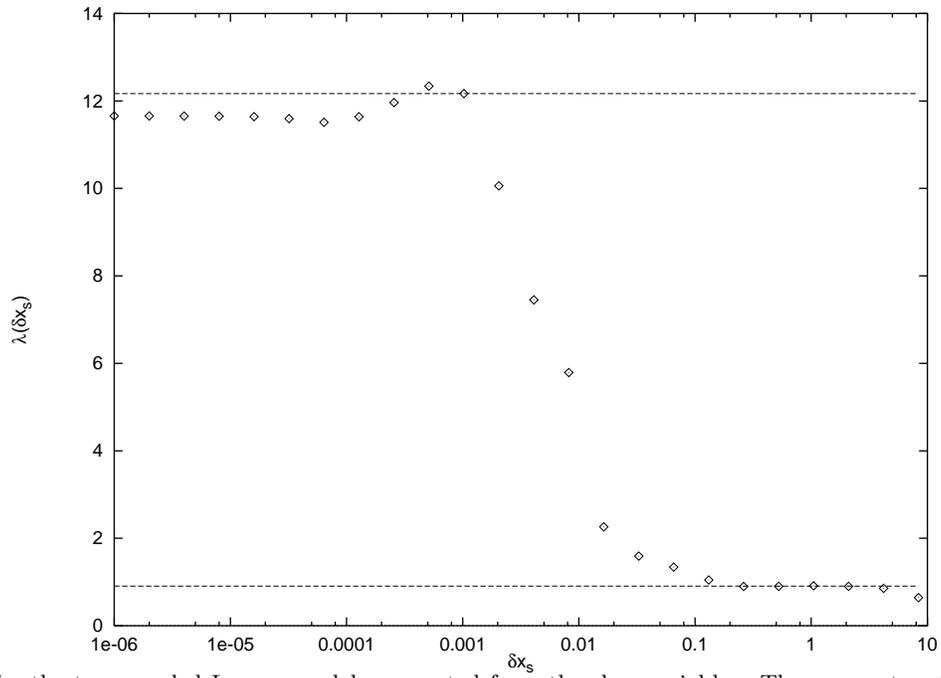}}
\caption{FSLE for the two coupled Lorenz models computed from the
slow variables. The parameters for the computation are:
$\delta_{0}=10^{-6}$, $m=25$, $r=2$ and $N=500$.
The two horizontal lines represent the uncoupled
Lyapunov exponents $\lambda^{(f)}$ and $\lambda^{(s)}$.
}
\label{fig3}
\end{figure}

We now apply the FSLE algorithm to the slow component of the the
error, $\delta x_{s}$ (figure \ref{fig3}).
We define a series of $m=25$
thresholds starting with $\delta_0=10^{-6}$ and ratio $r=2$. The results
presented (figure \ref{fig3}) are obtained after averaging over $N=500$ 
realizations. For very small $\delta$, the FSLE recovers the leading
Lyapunov exponent $\lambda_{max}$, indicating that in small scale
predictability the fast component has indeed a dominant role. As soon as the
error grows above the coupling $\epsilon_{s}$, $\lambda(\delta)$ drops
to a value close to $\lambda^{(s)}$ and the characteristic time of 
small scale dynamics is no more relevant.

\begin{figure}
\centerline{\psfig{file=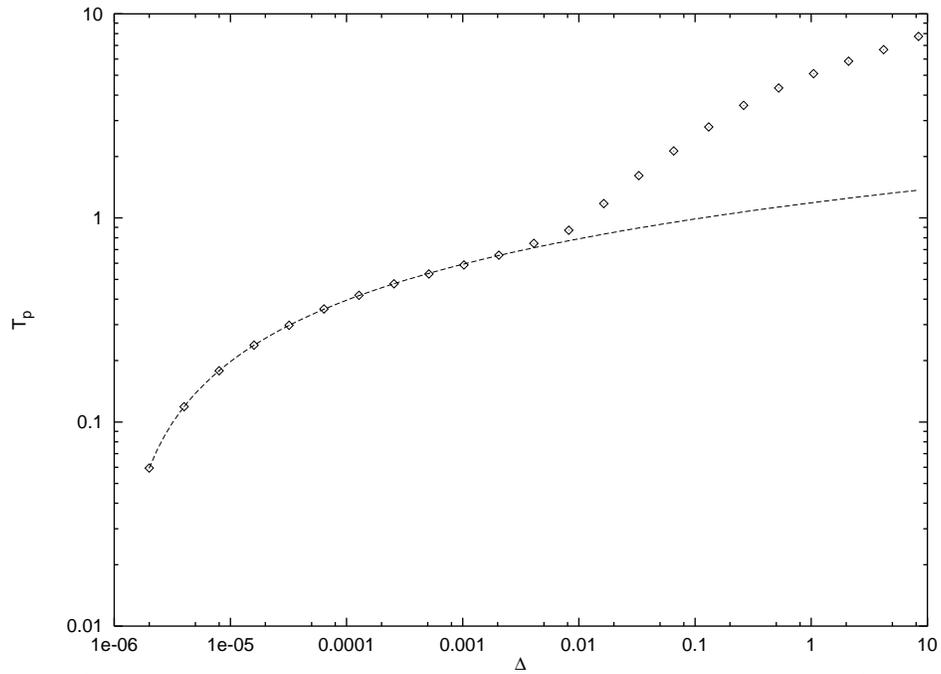}}
\caption{Predictability time for the slow component of the two coupled
Lorenz models as a function of the tolerance $\Delta$. The initial error
is fixed at $\delta=10^{-6}$. The dashed line represent the Lyapunov
estimation $T_{p}\sim \lambda^{-1} \ln(\Delta/\delta)$.
}
\label{fig4}
\end{figure}

In figure \ref{fig4} we plot the slow component predictability 
time (\ref{eq:b3}) for a fixed initial error $\delta x_{s}=10^{-6}$ 
as a function of the tolerance $\Delta$. We observe, as expected, an
enhancement of $T_{p}$ as soon as one accepts a tolerance larger than
the typical fast component fluctuation in the slow time series. Observe
that the naive application of (\ref{eq:1}) would heavily underestimate
the predictability time for large tolerance (dashed line).

\medskip

We now consider the second example. It is a more complex system
introduced by Lorenz 
(Lorenz, 1996) 
as a toy model for the Atmosphere dynamics which includes explicitly 
both large 
scales (synoptic scales, slow component) and small scales (convective 
scales, fast component). The apparent paradox described above can be 
reformulated here by saying that a more refined Atmosphere model (which
is able to capture the small scale dynamics) would be less predictable 
of a rougher one (which resolve only large scale motion) and thus the 
latter should be preferred for numerical weather forecasting. We will see 
that also in this case, the effect of the small, fast evolving, scales becomes 
irrelevant for the predictability of large scale motion if one consider
large errors.

The model introduces a set of large scale, slow evolving, variables 
$x_{k}$ and small scale, fast evolving, variables
$y_{j,k}$ with $k=1,...,K$ and $j=1,...,J$. As in (Lorenz, 1996)
we assume periodic boundary conditions on $k$ ($x_{K+k}=x_{k}$, 
$y_{j,K+k}=y_{j,k}$) while for $j$ we impose
$y_{J+j,k}=y_{j,k+1}$. The equation of motion are
\begin{equation}
\begin{array}{lll}
{\ts d x_{k} \over \ts dt} & = & - x_{k-1} \left(x_{k-2}-x_{k+1}\right)
- x_{k} + F - \sum_{j=1}^{J} y_{j,k} \\
{\ts d y_{j,k} \over \ts dt} & = & - c b y_{j+1,k} 
\left(y_{j+2,k}-y_{j-1,k}\right) - c y_{j,k} + x_{k} \\
\end{array}
\label{eq:c4}
\end{equation}
in which $c$ again represent the relative time scale between fast and slow
dynamics and $b$ is a parameter which controls the relative amplitude. 

Let us note that (\ref{eq:c4}) has the same qualitative structure of a finite
mode truncation of Navier-Stokes equation, with quadratic 
inertial terms and viscous dissipation. The coupling (with unit strength) is
chosen in order to have the ``energy'' 
\begin{equation}
E = {1 \over 2} \left( \sum_{k=1}^{K} x_{k}^2 + 
\sum_{k=1}^{K} \sum_{j=1}^{J} y_{j,k}^2 \right)
\label{eq:c5}
\end{equation}
conserved in the inviscid, unforced limit. The forcing term drives only
the large scales and we will consider $F=10$ which is sufficient for
developing chaos.

We have performed the computation of the FSLE for system (\ref{eq:c4})
with parameters as in 
(Lorenz, 1996): 
$K=36$, $J=10$, $c=b=10$ 
implying that the typical $y$ variable is 10 times faster and smaller
than the $x$ variable. In this case we choose to adopt for measuring the
errors the global Euclidean norm on both the slow and fast variables
(energy norm): this is for mimic a realistic situation in which we are 
not able to recognize {\it a priori} the slow component in the system.

\begin{figure}
\centerline{\psfig{file=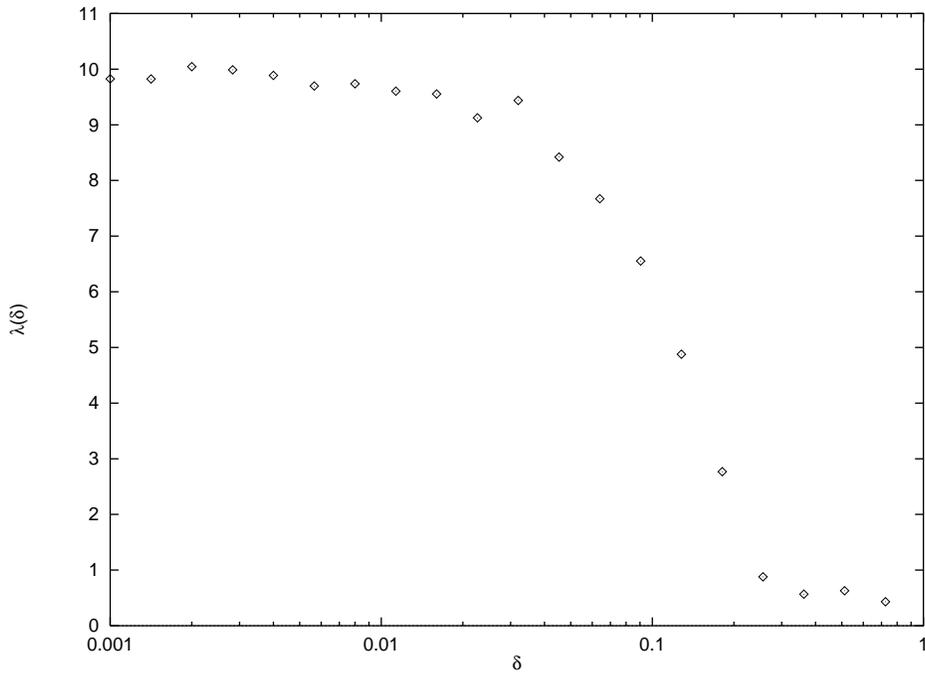}}
\caption{FSLE computed for the toy Atmospheric model.
The parameters for the computation are: $\delta_{0}=10^{-3}$, $m=20$,
$r=2^{1/2}$ and $N=1000$.
}
\label{fig5}
\end{figure}

The result of the FSLE computation is displayed in figure \ref{fig5} 
after averaging over
$N=1000$ realizations with initial error $\delta_{min}=10^{-5}$. We set
$m=20$ thresholds with $\delta_{0}=10^{-3}$ and ratio $r=2^{1/2}$.
For very small errors we
observe the saturation of $\lambda(\delta)$ to the leading Lyapunov
exponent of the system $\lambda_{max} \simeq 9.9$. For errors larger than the
typical r.m.s. value of the fast variables 
($\langle y^2 \rangle^{1/2}\simeq0.25$) we observe a second plateau
at $\lambda \simeq 0.5$, corresponding to the inverse characteristic time 
of large scales.
We observe that the relative time scale between fast and slow motions as
computed by the FSLE is slightly larger than the value of the parameter
$c$. We think that this effect is due to coupling which here
cannot be assumed small as in the previous example.

\begin{figure}
\centerline{\psfig{file=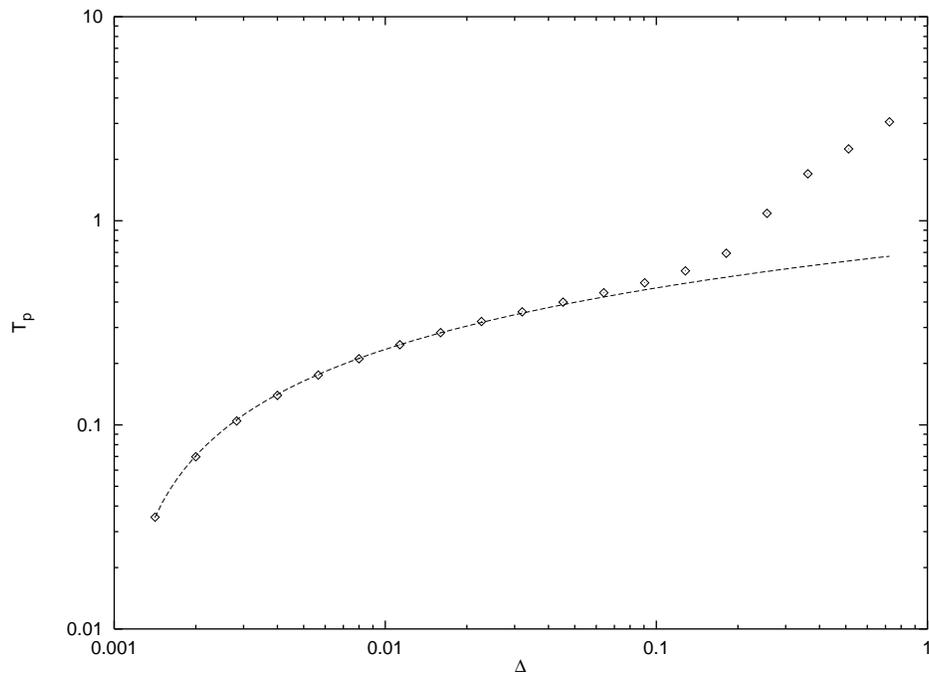}}
\caption{Predictability time for the toy Atmospheric model as a function
of the tolerance $\Delta$. The initial error is $\delta=10^{-3}$. The
dashed line represent the Lyapunov based estimation.
}
\label{fig6}
\end{figure}

In figure \ref{fig6} we plot the predictability time 
(\ref{eq:b3}) for fixed initial error $\delta=10^{-3}$ and
different thresholds. As in the previous example, we observe an 
enhancement of the predictability time for large tolerance $\Delta$ 
with respect to the Lyapunov exponent estimation. 
For large initial errors (as it is usually the case in numerical weather
forecasting) the predictability time is thus independent of the 
Lyapunov exponent.

\section{Conclusions}
\label{sec:4}
We have shown that in systems with possess different characteristic
time scales, the predictability time can be an independent quantity of
the leading Lyapunov exponent. The latter is usually associated to the
faster characteristic time and dominates the exponential growth of
infinitesimal errors. Large errors will evolve in general with 
large scale characteristic time which thus rules large scale
predictability.

We have introduced a generalization of the Lyapunov exponent which
allows to compute the average exponential error growth at a given error
size $\delta$. The Finite Size Lyapunov Exponent is expected to converge
at the leading Lyapunov exponent for very small errors. For larger
errors, $\lambda(\delta)$ is decreasing with $\delta$ and thus the FSLE
analysis predicts an enhancement of the predictability time as observed
in several numerical experiments.

We illustrate these concepts on two model examples which possess
different characteristic timescales. The numerical computation of the
FSLE confirms the predictability enhancement with respect to the
Lyapunov analysis.

Our results have a general significance which exceeds the 
proposed models. In particular, whenever one can identify in the
system different features with different intrinsic time scales, one
expects that slow varying quantities (i.e. large scale features) are
predictable longer than fast evolving quantities.
Moreover, our results demonstrate that the estimation of the
predictability time for a large scale circulation model do not
require to resolve the small scale dynamics.

\section*{Acknowledgments}
This paper stems from the work of Giovanni Paladin, who has been tragically 
unable to see its conclusion. We dedicate this paper to his memory.
G. Boffetta thanks the ``Istituto di Cosmogeofisica del CNR'', 
Torino, for hospitality and support. 
This work has been partially supported by the CNR
research project ``Climate Variability and Predictability''.




\end{document}